\documentclass[fleqn,10pt]{wlscirep}
\usepackage{graphicx}
\usepackage{subcaption}
\usepackage{amsmath}
\usepackage{amssymb}
\usepackage{float}
\usepackage{xcolor}
\title{Detecting Crystallographic Lattice Chirality using Resonant Inelastic X-ray Scattering}

\author[1]{Sean Mongan}
\author[2]{Zengye Huang}
\author[1,2,*]{Trinanjan Datta}
\author[3]{Takuji Nomura}
\author[2,*]{Dao-Xin Yao}
\affil[1]{Department of Chemistry and Physics, Augusta University, 1120 15$^{\text{th}}$ Street, Augusta, Georgia 30912, USA.}
\affil[2]{State Key Laboratory of Optoelectronic Materials and Technologies, School of Physics, Sun Yat-Sen University, Guangzhou 510275, China.}
\affil[3]{Synchrotron Radiation Research Center, National Institutes for Quantum and Radiological Science and Technology, SPring-8, 1-1-1 Kouto, Sayo, Hyogo 679-5148, Japan.} 
\affil[*]{tdatta@augusta.edu}
\affil[*]{yaodaox@mail.sysu.edu.cn}

\begin{abstract}
The control and detection of crystallographic chirality is an important and challenging scientific problem. Chirality has wide ranging implications from medical physics to cosmology including an intimate but subtle connection in magnetic systems, for example Mn$_{1-x}$Fe$_{x}$Si. X-ray diffraction techniques with resonant or polarized variations of the experimental setup are currently utilized to characterize lattice chirality. We demonstrate using theoretical calculations the feasibility of indirect $K$ -edge bimagnon resonant inelastic X-ray scattering (RIXS) spectrum as a viable experimental technique to distinguish crystallographic handedness. We apply spin wave theory to the recently discovered $\sqrt {5}\times\sqrt {5}$ vacancy ordered chalcogenide Rb$_{0.89}$Fe$_{1.58}$Se$_{2}$ for realistic X-ray experimental set up parameters (incoming energy, polarization, Bragg angle, and experimental resolution) to show that the computed RIXS spectrum is sensitive to the underlying handedness (right or left) of the lattice. A Flack parameter definition that incorporates the right- and left- chiral lattice RIXS response is introduced. It is shown that the RIXS response of the multiband magnon system RbFeSe arises both from inter- and intra- band scattering processes. The extinction or survival of these RIXS peaks are sensitive to the underlying chiral lattice orientation. This in turn allows for the identification of the two chiral lattice orientations.
\end{abstract}

\begin{document}

\flushbottom
\maketitle

\newcommand{\kvec}{\mathbf{\mathbf{k}}}
\newcommand{\ad}{a^{\dagger}}
\newcommand{\bd}{b^{\dagger}}
\newcommand{\g}{\gamma}
\newcommand{\de}{\delta}
\newcommand{\G}{\vec{\Delta}}
%
%
\thispagestyle{empty}

\section*{Introduction}
Chirality (right- or left- handedness) can arise in molecules and crystal structures both naturally and through synthesis \cite{barron2008}. Characterization and separation of enantiomorhpic (pair of chiral) crystals is a practical issue of utmost scientific importance. Examples of recent scientific studies where chirality was a crucial conceptual ingredient include the synthesis of chiral magnetic crystals Mn$_{1-x}$Fe$_{x}$Si relevant to the study of skyrmion (topological defects in spin texture) physics \cite{nakajimae1602562, griorievPhysRevB.81.012408}, nanotechnology applications of virus capsids \cite{zeng.acsnano.8b00069}, pharmacological action of synthetic drugs \cite{zhao2017}, and astrobiology \cite{feringa38.3418}. The effect of chirality is also the root cause behind magneto-chiral dichroism \cite{rikken1997}. Thus, an appropriate experimental technique to accurately determine the racemic conglomerate (an equimolar mixture of chiral pairs) composition can have wide ranging scientifc impact. 

Experimental techniques in chemistry have focused on asymmetric synthesis and its applications for both organic and inorganic molecules \cite{benmoshe2014,nandichiral,barron2008}. Controlling growth processes to ensure the synthesis of crystallographically chiral inorganic molecules can be difficult\cite{griorievPhysRevB.81.012408,hazePhysRevB.95.060415}. To develop methods of chirality control one needs an accurate tool to detect chirality composition. Some of the techniques to quantify racemic conglomerate composition or to determine the absolute structure include the chiroptical method\cite{ajm2002}, the Bijvoet method \cite{flacka22047}, anisotropic tensor susceptibility approach of Dmitrienko \cite{dmitrienkoa21876,dmitrienkoa23026}, and resonant circularly-polarized hard X-ray diffraction technique of Kousaka \emph{et. al.} \cite{kousakaJPSJ.78.123601,kousaka1742-6596-502-1-012019}.

The intensities of the Bijvoet pairs arising from reflections of Friedel opposites at $(h,k,l)$ and $(\bar{h},\bar{k},\bar{l})$ are inequivalent in a non-centrosymmetric chiral system. The current state-of-the-art equipment is capable of distinguishing intensity differences from Friedel opposites \cite{kousakaJPSJ.78.123601}. In this context, it is customary to define a Flack parameter which assists with the crystal structure analysis of a  noncentrosymmetric crystal arrangement from its inverse when the dominant scattering is anomalous\cite{flacksh0129}. However, crystal absorption effects typically limit the size of the sample that can be studied. A solution to this practical bottleneck (proposed by Dmitrienko) utilizes the anisotropic tensor susceptibility response to Bragg reflection difference between right-handed circularly polarized and left-handed circularly polarized X-ray beams \cite{dmitrienkoa21876,dmitrienkoa23026}. The experimental feasibility of this concept has been demonstrated by Tanaka \emph{et. al.}\cite{tanakaPhysRevLett.100.145502}. Improving on this Kousaka \emph{et. al.} showed that crystallographic chirality of CsCuCl$_{3}$ can be probed by resonant circularly polarized hard X-ray diffraction \cite{kousakaJPSJ.78.123601}.

\begin{figure}[H]
\centering
\includegraphics[width=1.0\linewidth]{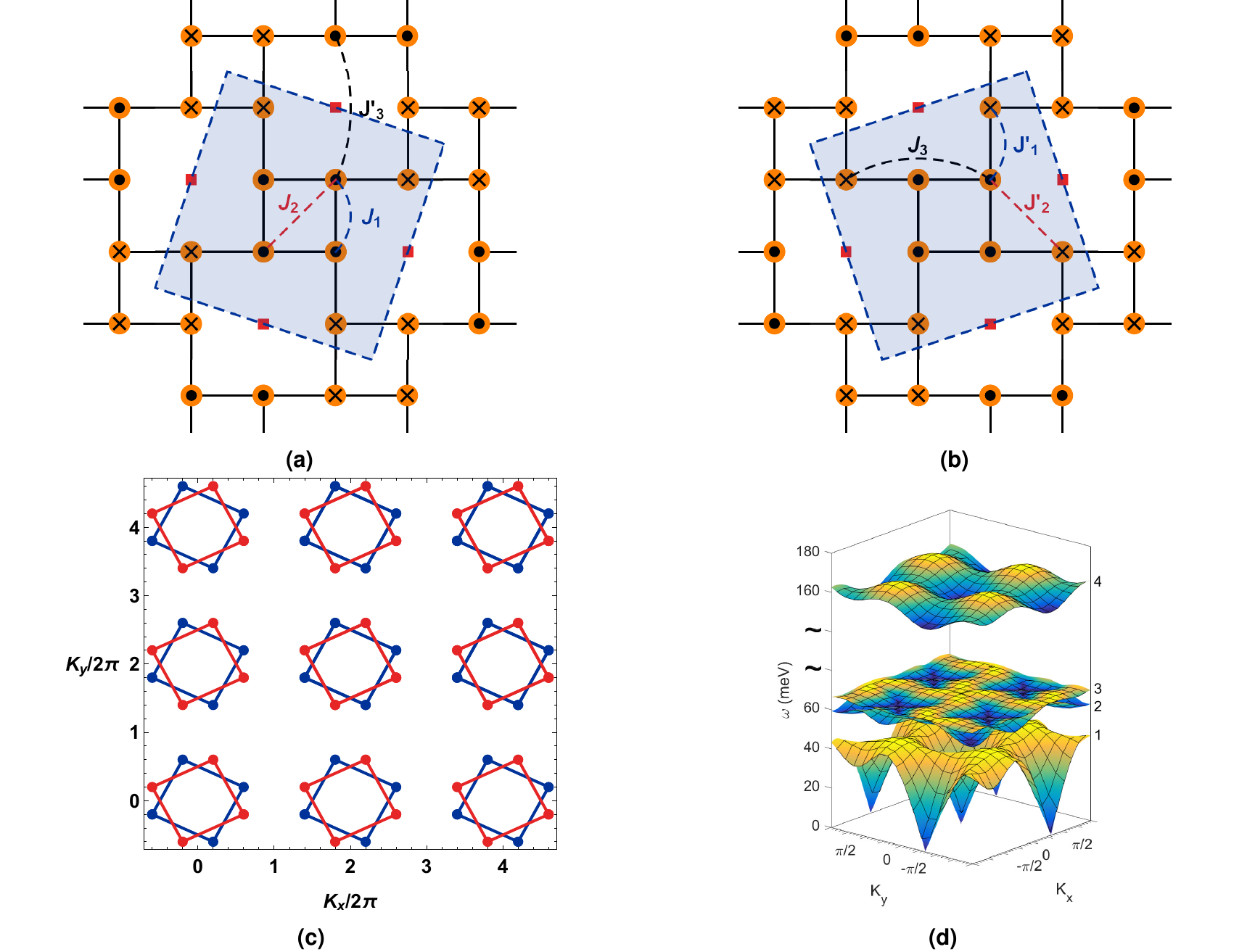}
\caption{Chiral lattice structures of Rb$_{0.89}$Fe$_{1.58}$Se$_{2}$ with the block antiferromagnetic spin ordering pattern. Exchange interactions, Brillouin zone momentum ordering wave vector combinations, and magnon energy bands are shown. Filled circles with solid dots represent up spins. Filled circles with crosses represent down spins. The shaded tilted squares are a guide to the two chiral lattice orientation. The exchange parameters (in meV) chosen for our calculation are $J_{1}=-30$, $J_{2}=20$, $J^{'}_{3}=0$ and $J^{'}_{1}=-10$, $J^{'}_{2}=20$, $J_{3}=9$, with spin $S=1$. (a) left chiral lattice. (b) right chiral lattice. (c) Bragg peak locations. The block antiferromagnetic checkerboard structure gives rise to eight antiferromagnetic Bragg peaks. The left chiral wave vector combinations are $(H_{o},K_{o},L_{o})$ = (0.2 + $m$,0.6 +$n$,$L_{o}$);(-0.2 + $m$,-0.6 +$n$,$L_{o}$);(0.6 + $m$,-0.2 +$n$,$L_{o}$);(-0.6 + $m$,0.2 +$n$,$L_{o}$). The right chiral wave vector combinations are $(H_{o},K_{o},L_{o})$ = (0.6 + $m$,0.2 +$n$,$L_{o}$);(-0.6 + $m$,-0.2 +$n$,$L_{o}$);(0.2 + $m$,-0.6 +$n$,$L_{o}$);(-0.2 + $m$,0.6 +$n$,$L_{o}$); $m,n = \{\pm 2, \pm 4, \hdots\}$. $L_{o}$ is set equal to zero since we do not consider the three dimensional model. (d) Magnon dispersion energy bands $\omega$ (meV). The bands are numbered one through four. The double wiggly lines represent a break in the energy axis since the the highest band is far above the lowest lying three bands. All the bands participate in the RIXS scattering process.}\label{fig:lattices}
\end{figure}

The study of skyrmion physics initiated the synthesis of chiral MnSi crystals. Surprisingly, synthesis of these crystals displayed a total chiral symmetry breaking \cite{nakajimae1602562}. Cyrstals of only one type of handedness (left), having $P2_{1}3$ symmetry, were observed even when grown from different seeds. This finding was unexpected since the related magnetic system Fe$_{1 -x}$Co$_{x}$Si with Dzyaloshinskii-Moriya interaction is known to exist in both its left- and right- handed configurations. The relationship between crystalline chirality and magnetic structure in Mn$_{1 -x}$Fe$_{x}$Si is thus subtle. The correlation between crystalline chirality and magnetic helicity is also different between the two compounds. In Fe$_{1 -x}$Co$_{x}$Si a right-handed magnetic helix coexists with a left-handed chiral lattice and vice versa. But, Grigoreiv \emph{et. al.}~\cite{griorievPhysRevB.81.012408} has experimentally shown that in MnSi the left (right)-handed lattice chirality exists with a left (right)-handed magnetic structure.

In the recently identified iron based superconductors chalcogenides form a category in which lattice chirality can be shown to exist in a right- and a left- handed form. The iron based superconductors (FeSC) belong to various families, with a classification scheme that includes five different families, namely - the 1111 system (RFeAsO, R = Lanthanide), the 122 system (AFe$_{2}$Se$_{2}$, A = Ba, Sr, Ca, K), the 111 system (AFeAs, A = Li, Na), the 11 system (Fe$_{1+y}$Te$_{1-x}$Se$_{x}$), and the alkali iron selenides (A$_{x}$Fe$_{2-y}$Se$_{2}$, A=K, Rb, Cs, Tl, ...). The last category includes the insulating vacancy ordered compound A$_{2}$Fe$_{4}$Se$_{5}$ (245 class) which is of interest for our calculation purpose due to its crystallographic chiral nature\cite{daiRevModPhys.87.855,dagottoRevModPhys.85.849,zhao2009,PhysRevB.84.155108,yangPhysRevB.94.024503,zavalijPhysRevB.83.132509}. The general consensus is that magnetic order in most of the parent compounds of FeSC are described by a bad metal description along with some variant (collinear, bicollinear-, block-, and block with $\sqrt{5}\times\sqrt{5}$ vacancy) of antiferromagnetic (AFM) ordering\cite{huPhysRevB.85.144403}. It has been experimentally validated that the chalcogenide Rb$_{0.89}$Fe$_{1.58}$Se$_{2}$, hereafter referred to as RbFeSe, is a vacancy ordered insulator in the undoped limt exhibiting a block-AFM structure with vacancy (BAFv), see Figure~\ref{fig:lattices} \cite{wang2011,wangPhysRevB.84.094504,yePhysRevLett.107.137003,chenPhysRevX.1.021020}. A local moment description of RbFeSe has provided a satisfactory understanding of its magnetic excitation and inelastic neutron scattering spectrum. The neutron scattering studies have revealed a Fe$_{4}$ block-AFM checkerboard (also known as a BAFv) structure. This block checkerboard orders with a $\sqrt{5}\times\sqrt{5}$ superlattice unit cell which can exist in either a left- or right- chiral orientation as shown in Figs.~\ref{fig:lattices}(a) and \ref{fig:lattices}(b)\cite{wang2011}. 

Experimental\cite{PhysRevLett.100.097001,PhysRevLett.102.167401,Nature.485.82,PhysRevLett.105.157006,PhysRevLett.108.177003,PhysRevB.85.214527,NatMater.11.850,clancyPhysRevB.95.235114}, theoretical\cite{amentRevModPhys.83.705,RevModPhys.73.203, nagaoPhysRevB.75.214414,brink0295-5075-80-4-47003,PhysRevB.77.134428,PhysRevB.86.125103,amentPhysRevLett.103.117003,Marra2016,perkinsPhysRevLett.117.127203}, and computational\cite{jiaPhysRevX.6.021020,PhysRevB.77.104519,kourtisPhysRevB.85.064423,nocera2018} investigations have demonstrated that resonant inelastic x-ray scattering (RIXS) spectroscopy can provide a rich source of physics information. Theoretical studies detailing the effects of magnetic frustration\cite{dattaPhysRevB.89.165103}, triangular lattice geometry\cite{dattaPhysRevB.92.035109}, magnon-phonon coupling\cite{dattaPhysRevB.96.144436}, and multi-band excitation\cite{dattaJPCM29.505802} of the bimagnon excitation at the $K$ -edge indirect RIXS have been reported. Motivated by the conceptual underpinnings of the aforementioned RIXS studies and the occurence of the chiral 245 class of chalcogenides we investigate the following question -- ``Can indirect $K$ -edge RIXS spectrosocpy assist with the detection of lattice chirality?" The scientific importance of a spectroscopic approach to analyze lattice chirality has been highlighted earlier. Within the context of the present article a new chirality detection method will benefit several scientific communities in physics including RIXS spectroscopy and highly frustrated quantum magnetism. Using RbFeSe as a realistic model system, we conclude from our calculations that indirect $K$ -edge RIXS for bimagnons is a viable spectroscopic method to distinguish lattice chirality. Our theory considers realistic experimental conditions that include the effect of incoming X-ray polarization and Bragg angle effects (see Fig. \ref{fig:rlchiral}a and supplementary Fig. S1 available online) and experimental resolution effects (see Fig. \ref{fig:RMixROWVmn24full}). We theoretically investigate various experimental set-ups to propose the optimal scattering condition for RIXS intensity detection\cite{hancockPhysRevB.82.035113}. 

The starting point of our theoretical analysis is a highly frustrated J$_{1}$-J$_{2}$-J$_{3}$ quantum Heisenberg Hamiltonian. This magnetic model provides a universal description of the local moment description of the 122 iron chalcogenides such as A$_{y}$Fe$_{2-x}$Se$_{2}$. Energy comparison studies have revealed that the vacancy-ordered states are energetically favored for the reduction of magnetic frustration. The BAFv phase can be modeled using six distinct interaction coefficients $J_{ij}\in \{J_1,J_1',J_2,J_2',J_3,J_3'\}$, see Fig.~\ref{fig:lattices} caption for exchange parameter values~\cite{fangPhysRevB.85.134406}. The Hamiltonian is written as
\begin{eqnarray}
\label{eq:hamrfs}
    H&=&J_{1}\sum_{i, \de, \de '(>\de)} \mathbf{\mathbf{S}}_{i,\de}\cdot\mathbf{\mathbf{S}}_{i,\de '}+J_{1}'\sum_{i, \de, \de ',\gamma} \mathbf{\mathbf{S}}_{i,\de}\cdot\mathbf{\mathbf{S}}_{i+\g,\de '} +J_{2}\sum_{i, \de, \de^{\prime \prime}(>\de)} \mathbf{\mathbf{S}}_{i,\de}\cdot\mathbf{\mathbf{S}}_{i,\de^{\prime \prime}}+J_{2}'\sum_{i, \de, \de^{\prime \prime},\gamma} \mathbf{\mathbf{S}}_{i,\de}\cdot\mathbf{\mathbf{S}}_{i+\g,\de^{\prime \prime}}\nonumber\\
    &+&J_{3}\sum_{i, \de, \de^{\prime \prime \prime},\gamma} \mathbf{\mathbf{S}}_{i,\de}\cdot\mathbf{\mathbf{S}}_{i+\g,\de^{\prime \prime \prime}}+J_{3}'\sum_{i, \de, \de^{\prime \prime \prime},\gamma'} \mathbf{\mathbf{S}}_{i,\de}\cdot\mathbf{\mathbf{S}}_{i+\g',\de^{\prime \prime \prime}},
\end{eqnarray}
where the index $i$ ranges over all of the spin blocks in the system. The spin $\mathbf{S}_{i,\de}$ is the $\de$-th spin in the $i$-th block. The lattice coordination vector $\de'$ ranges over the number of the spin sites that are in nearest-neighbor (NN) positions to the $\de$-th site, $\de^{\prime \prime}$ ranges over the number of the spin sites that have next-nearest-neighbor (n-NN) interactions with $\de$-th spin site, and $\de^{\prime \prime \prime}$ ranges over the number of the spin sites that have next-next-nearest-neighbor (nn-NN) interactions, $\g$ ranges over the number of the block sites that have NN interactions to the $i$-th block, and $\g'$ ranges over the number of block sites that have the n-NN interaction to the $i$-th block. We will use this model to compute the indirect $K$ -edge bimagnon RIXS spectrum to highlight the effects of lattice chirality detection.

\begin{figure}[h]
    \centering
\includegraphics[width=0.85\linewidth]{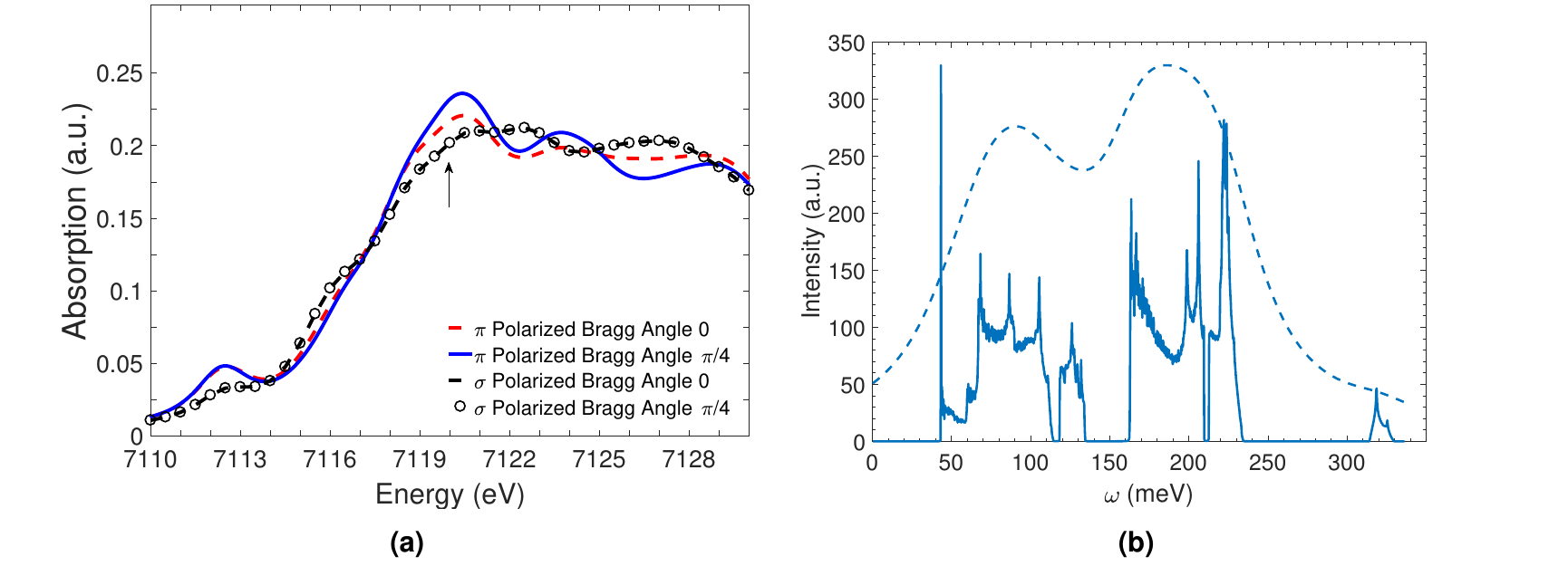}
    \caption{X-ray absorption and multiband bimagnon RIXS spectrum. (a) Calculated x-ray absorption spectra (XAS) of Rb$_{0.89}$Fe$_{1.58}$Se$_{2}$ for various polarization and scattering geometries. $\pi$ [$\sigma$] polarization means that both the polarization directions of incoming and outgoing x-rays are parallel [perpendicular] to the scattering plane. (b) RIXS plot without the resonance factor dependence and experimental resolution for the right chiral lattice with a $\{2,4\}$ momentum combination. Intra- and inter- band transitions give rise to multiple peaks.}\label{fig:xasrixs}
\end{figure}

The scattering cross section between hard x-rays and matter can be increased 
by taking advantage of the resonance absorption. In our calculation, we consider the Fe $K$ -edge.
To obtain sizable RIXS intensity, we need to tune the incident x-ray energy 
to an absorption energy characteristic of the constituent elements. 
The x-ray absorption spectra (XAS) at the $K$-shell resonance can be approximately calculated from the density of states (DOS) 
of the conduction $p$ states $\rho_{\ell}(\varepsilon)$ ($\ell=x,y,z$) above the Fermi level ($E_F$) 
\begin{equation}
I_{XAS}({\bf e}_i,\omega_i) = - 2 \sum_{\ell} |w_{\ell}({\bf e}_i)|^2 \int_{E_F}^{\infty} 
\frac{d \varepsilon}{\pi} 
\Im m \biggl[ \frac{\rho_{\ell}(\varepsilon) }{\gamma(\omega_i; \varepsilon)} \biggr],  
\end{equation}
where $\rho_{\ell}(\varepsilon)$ is the Fe-$p_{\ell}$ DOS, 
$\gamma(\omega; \varepsilon) = \omega + \varepsilon_K + i \Gamma_K - \varepsilon$ 
with $\varepsilon_K$ and $\Gamma_K$ being the energy and damping of the $K$ shell level, 
and $w_{\ell}(\bf{e})$ is the electric-dipole transition matrix for x-ray polarization ${\bf e}$, 
given by 
\begin{equation}
w_{\ell}({\bf e}) = - \frac{e}{m}\sqrt{\frac{2\pi}{\omega_{i}}} {\bf e} 
\cdot \langle p_{\ell} |{\bf p}|1s \rangle \propto {\bf e} \cdot {\bf e}_\ell, 
\end{equation}
where the elementary charge $e$ and the electron mass $m$ are expressed in natural units ($c = \hbar = 1$). 
To obtain the $K$ -edge XAS for RbFeSe, we calculate Fe-$p$ DOS $\rho_{\ell}(\varepsilon)$ using the WIEN2k code~\cite{wien2k:2001}. 
In Fig.~\ref{fig:xasrixs}a, we present the calculated results of XAS for several scattering geometries where we set $\varepsilon_K = -7112$ eV and $\Gamma_K = 1$ eV in our numerical calculation. 
$\Gamma_K$ determines the life time of the core hole created in the intermediate state of RIXS,
$\hbar/\Gamma_K \sim $ femtoseconds. 

The indirect bimagnon RIXS amplitude can be expressed by the product of the bimagnon-excitation part 
and a resonance factor part within the lowest-order approximation in the coupling between the core-hole and magnons 
~\cite{nomuraPhysRevB.96.165128,brink0295-5075-80-4-47003,nagaoPhysRevB.75.214414}. The bimagnon-excitation part is given by 
\begin{equation}
\label{eq:o2mb1}
    \mathcal{O}_{2}(\mathbf{\mathbf{q}})= 
\sum_{\kvec} \sum_{i,j} \left( W_{ij}b^{\dagger}_{i \kvec+\mathbf{\mathbf{q}}}b^{\dagger}_{j -\kvec}
+ W_{ij}^*b_{i \kvec+\mathbf{\mathbf{q}}}b_{j -\kvec}\right),
\end{equation}
where $\mathbf{q}$ is the RIXS transfer momentum, the matrix elements $W_{ij}$ are defined in equations (\ref{eq:orixsmat1}) and (\ref{eq:orixsmat2}), $i,j = 1, 2, 3, 4$  are band indices, $b^{\dagger}_{i\kvec} (b_{i\kvec})$ represent creation (annihilation) operators for band index $i$ and momentum vector $\kvec$, respectively. Note, within the UCL expansion scheme the final RIXS operator expression is similar to equation (\ref{eq:o2mb1}) which is essentially the Born approximation result in the weak coupling limit~\cite{brink0295-5075-80-4-47003}. The resonance factor contribution is given by 
\begin{eqnarray}
R({\bf e}_i, {\bf e}_o; \omega_i; \omega) = 2 V \sum_{{\ell}=x,y,z} 
w_{\ell}({\bf e}_i) w_{\ell}^*({\bf e}_o) 
\int_{E_F}^{\infty} d\varepsilon 
\frac{\rho_{\ell}(\varepsilon)}{\gamma(\omega_i; \varepsilon) \gamma(\omega_o; \varepsilon)}, 
\end{eqnarray}
where the coefficient $V$ denotes the local coupling between the core hole and a pair of magnons. 
${\bf e}_i$ [${\bf e}_o$] and  $\omega_i$ [$\omega_o = \omega_i - \omega$] 
represent the polarization and energy of the incoming [outgoing] x-ray, respectively. 
When the incoming x-ray energy $\omega_i$ is tuned near the resonance absorption energy, 
then $|\gamma(\omega_i; \varepsilon)|$ is small, and the resonance factor as well as XAS intensity 
becomes significantly enhanced. The full RIXS intensity is expressed by the product of the resonance factor part 
and the bimagnon correlation function
\begin{equation}
I({\bf q}, \omega) = |R({\bf e}_i, {\bf e}_o; \omega_i; \omega) |^2\mathsf{S}({\bf q},\omega), 
\end{equation} 
where $\mathsf{S}({\bf q},\omega)$ is the bimagnon correlation function computed by means of the spin-wave theory as 
\begin{equation}
\label{eq:rixsmulb}
   \mathsf{S}({\bf q},\omega)= -\frac{1}{\pi}\Im m\left[\sum\limits_{\kvec}\sum_{ij}|W_{ij}|^2\frac{1}{\omega-\omega_{i,\kvec+\mathbf{\mathbf{q}}}-\omega_{j-\kvec}+i0^+}\right]. 
\end{equation}
Here we should note that, in tetragonal cases as FeSC, the resonance factor as well as XAS does not depend on the in-plane azimuthal angle 
and furthermore on chirality, since $\rho_x(\varepsilon)=\rho_y(\varepsilon)$. 
In all our subsequent RIXS calculations we choose 7120 eV (indicated by the arrow in Fig.~\ref{fig:xasrixs}a) as the incoming 
x-ray energy choice to maximize the RIXS intensity. 
\section*{Results}
In Fig.~\ref{fig:xasrixs}b we show the results of the bimagnon RIXS spectrum without the resonance factor and considering (for this discussion) zero experimental resolution. In contrast to the analysis of a simple two dimensional square lattice with spin $1/2$, which has a single magnon band, for RbFeSe we need to consider all possible intraband and interband RIXS transitions. Figure~\ref{fig:lattices}d shows all the four bands participating in the RIXS scattering process. 
\begin{figure}[H]
    \centering
\includegraphics[width=0.95\linewidth]{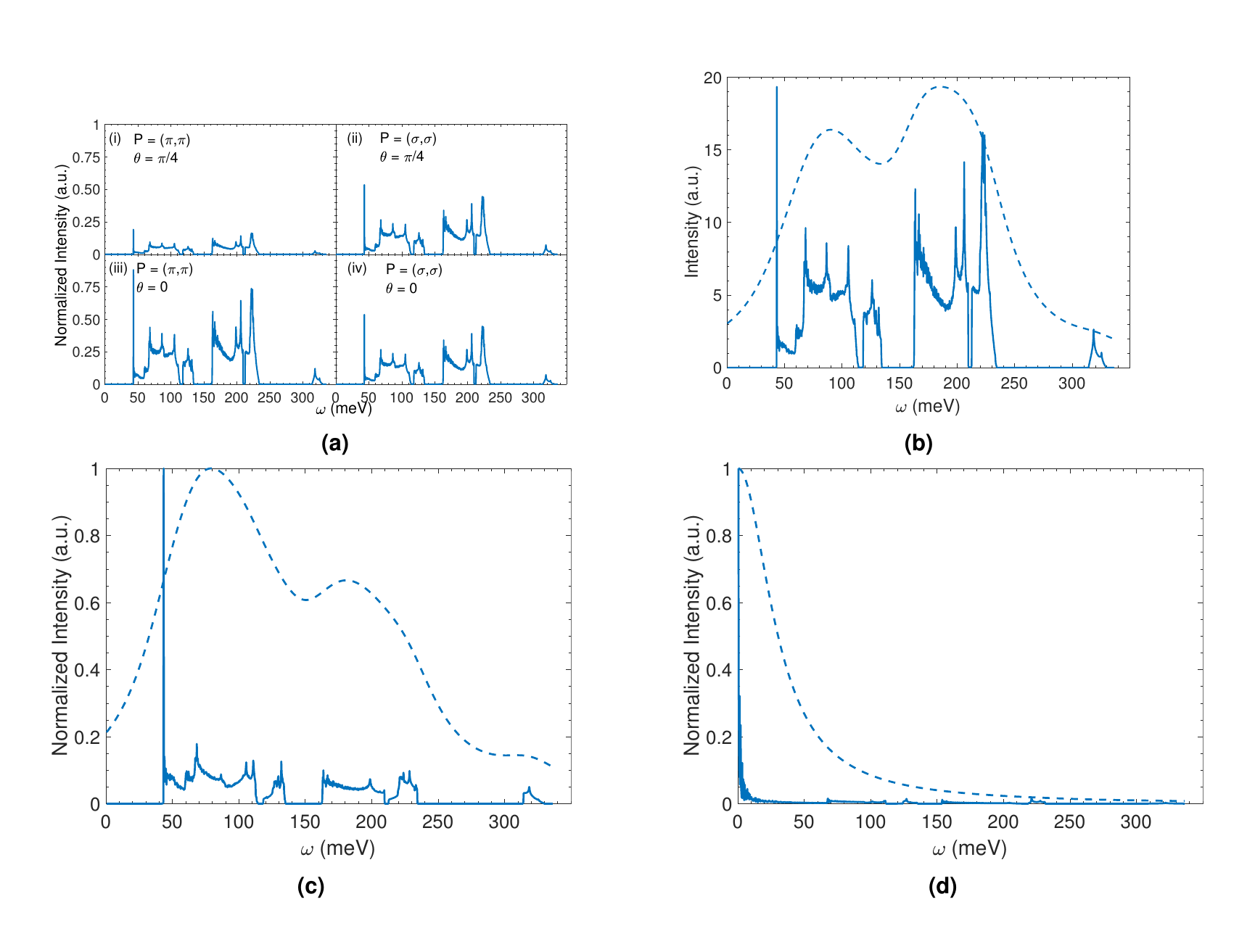}
    \caption{7120 eV Fe K -edge bimangon indirect RIXS response of the chiral right and left lattices of Rb$_{0.89}$Fe$_{1.58}$Se$_{2}$.  The feasibility of utilizing bimagnon RIXS spectroscopy at the K -edge to distinguish and detect lattice chirality is explicitly demonstrated. Dashed line represents 30 meV experimental resolution convoluted RIXS response. (a) Polarization (P) and Bragg angle $(\theta)$ dependence of the RC ROWV response. (b) RC ROWV response with resonance factor contribution for Bragg angle choice of zero and $(\pi,\pi)$ polarization. (c) LC LOWV response. (d) LC ROWV response. RC (LC) stands for right (left) chiral. ROWV (LOWV) stands for right (left) ordering wave vector. The Bragg peak momentum combination choice for all the calculations was $\{2,4\}$.}\label{fig:rlchiral}
\end{figure}
This multiband structure causes the RIXS operator to assume a matrix form, see equation (\ref{eq:orixsmat2}). Prior work by two of the authors (T.D. and D.X.Y) has highlighted the intricacies of multiband RIXS scattering processes giving rise to multipeak structures\cite{dattaJPCM29.505802}. The response of the right chiral lattice with the ROWV can be separated into three energy ranges that lie between 40 - 135 meV, 160 - 235 meV, and 160 - 330 meV. In the absence of experimental resolution multiple intra- and inter-band peaks are possible. Naturally, with the inclusion of experimental resolution most of these peaks are washed away. Please see supplementary Fig. S2 and Fig. S3, both available online, for assistance with associating the RIXS peaks with the corresponding band transition. All our calculations were performed for the $(2,4)$ momentum combination, see Figure~\ref{fig:lattices}c.

The checkerboard lattice supports acoustic spin waves which arises from eight Bragg peaks. We have verified that irrespective of the momentum combination choice, see Fig.~\ref{fig:lattices}c, the RIXS response is the same. We also infer from our investigation that at any location along the Bragg peak locations formed by connecting the right-ordering-wave-vector (ROWV) points on the right chiral (RC) lattice the RIXS response is the same. We find from our calculations that when the lattice chirality matches up with the probe ordering wave vector, for example right chiral with ROWV, there is a multi-peak contribution with two major RIXS peaks and one minor one. One of these is centered on 90 meV. The other is at 190 meV. There is a high energy sub-dominant intraband contribution above 300 meV arising from the 4-4 scattering channel. The peak at 190 meV slightly dominates in intensity compared to the lower 90 meV peak. The former arises from scattering process between the fourth band and the rest of the energy bands, i.e., 1-4, 4-1, 2-4, 4-2, 3-4, and 4-3. The remaining channels contribute to the 90 meV peak.

In order to make our results as quantitatively realistic as possible with experiments we take full account of the effects of the resonance factor described earlier and a 30 meV experimental resolution. We incorporate the resonance enhancement arising from XAS by tuning the incident x-ray energy around the maximum of XAS, i.e. around the main absorption energy 7120 eV. The results are displayed in Figure~\ref{fig:rlchiral}a. In general the resonance factor in $K$-edge RIXS does not bring about any significant momentum dependence into the total RIXS intensity. As a consequence momentum dependence of RIXS spectra is attributable to that of the corresponding electron correlation function. It appears that for $\sigma$ polarization the Bragg angle set up does not have a significant effect, but the $\pi$ polarization is sensitive to the Bragg scattering angle. Henceforth, we use this polarization and angle combination for our calculations. In Fig.~\ref{fig:rlchiral}b we display the results of the resonance factor, both with and without experimental resolution. Comparing the spectrum with Fig.~\ref{fig:xasrixs}b we find that the effect of the resonance factor is to cause a reduction in the intensity. The momentum dependence is unaffected, as expected. From an experimental perspective the $(\pi,\pi)$ beam polarization with a Bragg angle of zero maximizes the intensity. Inclusion of the experimental resolution removes the sharp multiple peaks in favor of a two-peak enveloping spectrum. Our recommendation to the experimentalists is to use this set up for an experimental verification of the proposed theory. 

In Figs.~\ref{fig:rlchiral}c and ~\ref{fig:rlchiral}d we display the RIXS response of an X-ray photon at the right- or left- ordering-wave-vector on a LC lattice. The response of the left chiral lattice with the LOWV is similar in its energy range with the right chiral lattice with the ROWV. However, these two spectra are not identical. In fact, for the left chiral lattice the 90 meV peak is more intense than the 190 meV location. When the lattice chirality is mismatched with the probe ordering wave vector, all the intra- and inter-band transitions are substantially muted in intensity in comparison to the 1-1 transition. This is primarily an elastic peak contribution which is absent when the chirality matches up with the ordering wave vector. The 1-1 channel is singular because of a vanishing denominator of $\Pi^{0}_{ij}(\mathbf{\mathbf{q}},t;\kvec,\kvec')$ where $\omega=\omega_{1,\kvec+\mathbf{\mathbf{q}}}=\omega_{1-\kvec}= 0$. These results hold considering realistic experimental resolution into account.

The response of the LC lattice interacting with the right (left) -ordering-wave vector produces a strikingly different signal. We notice that when the lattice chirality matches up with the incoming ordering wave vector chirality the system preserves its multipeak structure. In this example, it could be the multiple sharp peaks (at zero resolution), or the two-peak structure at finite experimental resolution. But, when it is mismatched a single dominant elastic peak arises. We should note that the other channels of RIXS scattering are not entirely extinguished. This difference in the responses (including resolution or not) allow us to conclude that the $K$ -edge RIXS signal is sensitive to lattice chirality. Thus we have demonstrated the feasibility of our claim that $K$ -edge bimagnon RIXS signal can be used to differentiate and identify lattice chirality. A point of concern is the presence of the strong elastic peak signal for wave vector lattice chirality mismatch where our predicted signal may be hiding. However, note that after the background elastic peak subtraction we expect a negligible RIXS signal from the rest of the intra- or interband scattering channels. Thus, our proposal to test lattice chirality will still hold. In fact, since substantial difference between left and right chiral lattices appears already at the elastic intensity $\omega=0$, elastic scattering (RXES) could also suffice to distinguish the chiral lattices. However, a knowledge of the inelastic part $(\omega>0)$ can lend insight into the potential multiband scattering processes.

\begin{figure}[h]
    \centering
\includegraphics[width=1.0\linewidth]{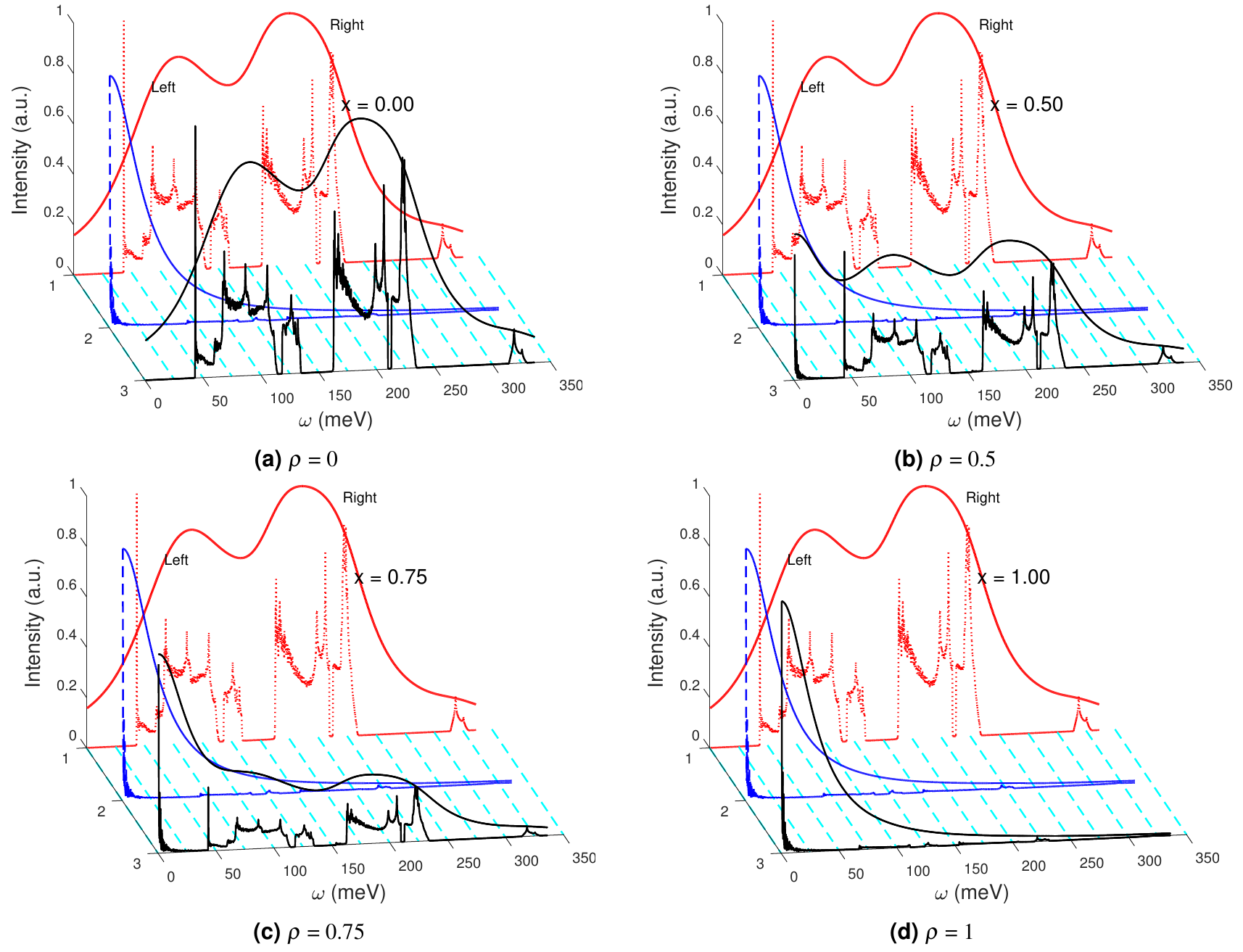}
\caption{Chiral composition mixture response for Flack parameter $\rho$ variation with 30 meV resolution. All responses are for the right chiral ordering wave vector and includes the effects of the resonance factor. (a) $\rho=0.0$ (pure right chiral), (b) $\rho=0.5$ (racemic conglomerate), (c) $\rho=0.75$, and (d) $\rho=1.0$ (pure left chiral).}
\label{fig:RMixROWVmn24full}
\end{figure}
\section*{Discussion}
In this article we report our findings on the detection of lattice chirality using $K$ -edge bimagnon indirect RIXS spectroscopy. Radiation incident at right or left ordering wave vector leads to distinct RIXS signals. Based upon the type of the ordering wave vector and the underlying lattice chirality the multipeak structures arising from intra- and inter- band transitions either survive or are almost extinguished (on a scale relative to the dominant peak). The extinction or survival of these RIXS peaks allows for the identification of the two chiral lattice orientations. To analyze experimental data that can include both types of chiral lattices we introduce the chiral Flack parameter $\rho$ as
\begin{equation}
\label{eq:flack}
 I({\bf q}, \omega) = (1-\rho)~[I({\bf q}, \omega)]_{\text{rch}} + \rho[I({\bf q}, \omega)]_{\text{lch}},
\end{equation}
where $[I({\bf q}, \omega)]_{\text{rch (lch)}}$ stands for right (left) chiral lattice RIXS intensity contribution. 

In Fig.~\ref{fig:RMixROWVmn24full} we track the evolution of the RIXS peaks of a chiral mixture of left- and right- lattices as the Flack parameter $\rho$ is varied. We find that our chirality detection proposal holds true irrespective of whether experimental resolution is included or not. We have tested the validity of our proposal over additional resolution values (10 meV, 15 meV, 50 meV, and 70 meV), see Supplementary Figs. S4 - S7. Based on our calculations we conclude that even beyond the 50 meV mark, detection of lattice chirality is still feasible (within the context of RbFeSe). Additionally, it is possible that a quasi-elastic line can possibly hide the magnetic excitation weight. We expect, in real experiments, appropriate subtraction of such a quasi-elastic line may be necessary still in the current resolution limit of 30-50 meV. According to the calculated results, the tail of the quasi-elastic line should be within less than 100meV, roughly speaking, which is about the excitation energy of the first main peak. On the other hand, we believe that further advancements in experimental resolution coupled with instrumentation techniques which can suppress the quasi-elastic tail will make the low-energy magnetic excitations observable. We also stress that although we chose RbFeSe as one example among various chiral materials, our calculations are not restricted only to Fe compounds. Our analysis approach can be generalized to other chiral transition-metal compounds. Furthermore, specific to the case of RbFeSe RIXS at the $K$ -edge, magnetic excitations can lie on top of dominant fluorescent peaks. But, these fluorescence peaks will shift along the energy loss axis, depending on the incident x-ray energy, while the RIXS weight will maintain the same energy-loss position due to its Raman-like character. Therefore, it is still possible to observe low-energy magnetic excitations even when the fluorescence peaks vanish or survive.

To establish an experimental protocol we suggest the following steps. First, perform the RIXS experiment using either the right- or left- chiral ordering wave vector. If the signal produces a single dominant peak then the lattice is of opposite chirality. If a multipeak response survives then the chirality is the same as that of the ordering wave vector. The only caveat here is to ensure that the data allow for each and every peak to be recorded up to its full intensity. Otherwise, there is a possiblity of misidentifying low intensity responses as a multipeak response. Using the signal from the right- or left- ordering wave vector as a baseline and by comparing the appearance and disappearance of the RIXS peaks one can deduce the chiral Flack parameter composition number $\rho$. For systems with more complicated unit cells multiple magnon bands are possible. While it maybe difficult to pinpoint the exact channel contributions to the peaks since the number of bands $\mathcal{N}$ causes the RIXS channels to increase as $\mathcal{N}^{2}$, the basic premise of the technique would still hold.
 
We have tested the validity of our prediction successfully for other commonly reported magnetic couplings relevant to the RbFeSe J$_{1}$-J$_{2}$-J$_{3}$ model~\cite{wang2011,fangPhysRevB.85.134406}. Additionally, the $S=1$ spin value will have reduced quantum spin fluctuation effects compared to $S=1/2$ where higher order interaction corrections are required\cite{dattaPhysRevB.89.165103}. Thus, the overall qualitative shape of the predicted RIXS spectrum will survive. The feasbility of the proposed approach is not restricted to collinear magnetic systems. In principle, even for non-collinear magnets that can display lattice chirality and support stable bimagnon excitations it should be possible to distinguish the chiral lattices using our approach. However, at present it is unclear what would be the consequences for a material where both the lattice and the magnetic chiralities are strongly coupled. 
It is possible that neutron scattering could offer an alternate way to detect lattice chirality, at least magnetic chirality, since the difference is evident in the one-magnon dispersion level~\cite{moonPhysRev.181.920}. But, RIXS has a practical advantage due to its small sample size requirement compared to neutron scattering. For example, it is difficult to grow samples of sizeable volume for the 1111 class of iron-pnictides. 

Two-magnon RIXS is possible also at ligand $K$ -edge in the soft x-ray regime. For copper oxides, two-magnon excitations and their dispersion have been observed at the O $K$ -edge\cite{PhysRevB.85.214527,ishiiPhysRevB.96.115148}, although the range of momentum transfer is somewhat restricted. Within the theoretical framework of this article, the two-magnon RIXS at ligand $K$ -edges can be treated in a similar way as in the cases of transition-metal $K$ -edges. We expect the only difference to be in the microscopic process to derive the magnon-core-hole coupling. The typical beamline resolutions are currently 30 meV for both $K$- and $L$ -edges. For the Cu -$K$ edge 25 meV has been achieved at DESY\cite{ketenogluhf5287}, but, for other transition-metal K -edges, there are no beamlines higher than 100 meV, to our full knowledge. Generally, intensity of $K$ -edge spectra becomes too weak as the resolution is increased.  While resolution issues still plague the $K$ -edge (compared to the $L$ -edge), the potential to utilize a multimagnon excitation to unravel lattice chirality effect offers an exciting future for RIXS spectroscopy in general. Finally, we hope our results will provide X-ray beamline scientists and experimentalists, engaged in RIXS research, with an impetus to further the limits of $K$ -edge bimagnon detection and its potential lattice chirality application.

\section*{Methods}
\subsection*{Spin wave Hamiltonian}
The Hamiltonian, equation (\ref{eq:hamrfs}), was treated using the linear spin wave theory (LSWT) approach. Using the standard Holstein--Primakoff transformation each spin operator $\mathbf{S}_i$ was transformed as $S_i^+=\sqrt{2S}a_i$, $S_i^-=\sqrt{2S}\ad_i$, and $S_i^z=S-\ad_ia_i$, where $a_{i}$ ($a^{\dagger}_{i}$) represents the bosonic annhilation (creation) operators at site $i$, respectively. The bosonic basis was Fourier transformed to $a_i=\frac{1}{\sqrt{N}}\sum_{\kvec}e^{i\kvec\cdot \mathbf{\mathbf{R}}_i}a_{\kvec}$ and $\ad_i=\frac{1}{\sqrt{N}}\sum_{\kvec}e^{-i\kvec\cdot \mathbf{\mathbf{R}}_i}\ad_{\kvec}$, where $R_i$ is the position of site $i$, $N$ is the number of magnetic unit cells, and $a_{\kvec}$ and $a^{\dagger}_{\kvec}$ are the momentum dependent bosonic operators. After performing the LSWT transformation the Hamiltonian can then be written as $H = E_0+\sum_{\kvec}\psi^{\dagger}_{\kvec}\mathcal{H}_{\kvec}\psi_{\kvec}+O(1/S)$, where $E_0$ is the classical ground state energy of the system, $\psi_{\kvec}^{\dagger}=[\ad_{\kvec 1}\text{ } \ad_{\kvec 2}\text{ } \ad_{\kvec 3}\text{ } \ad_{\kvec 4}\text{ } a_{-\kvec 1}\text{ } a_{-\kvec 2}\text{ } a_{-\kvec 3}\text{ } a_{-\kvec 4}]$ is the basis set, $\mathcal{H}_{\kvec}$ is Hamiltonian matrix, and $O(1/S)$ represents the non-linear spin wave theory terms. A paraunitary diagonalization was performed to calculate the energy modes of the bosonic excitations, see Figure~\ref{fig:lattices}d. At a given momentum $\kvec$ the energies of the bosonic modes are given by the eigenvalues to the expression $I_p\mathcal{H}_{\kvec}$ where 
\begin{eqnarray}
I_p=
    \begin{bmatrix}
    I & 0\\
    0 & -I\\
    \end{bmatrix}
\end{eqnarray}
is the paraunitary matrix and $I$ and $0$ are the identity and null matrices of dimension $4 \times 4$, respectively. The eigenvector solutions $\mathbf{X}_{\kvec n}$ to 
\begin{equation}
    I_p\mathcal{H}_{\kvec}\mathbf{X}_{\kvec n}=\lambda_{\kvec n}\mathbf{X}_{\kvec n} \label{Eq: eig}
\end{equation} 
are the Bogoliubov transformation coefficients for the n--th energy eigenmode $\lambda_{\kvec n}$. The Bogoliubov transformed basis is given by $\psi'_{\kvec}=\mathbf{X}_{\kvec}^{-1}\psi_{\kvec}$, where $\mathbf{X}_{\kvec}=[\mathbf{X}_{\kvec 1}\text{ } \mathbf{X}_{\kvec 2} \text{ } \mathbf{X}_{\kvec 3} \mathbf{X}_{\kvec 4}\text{ } (I_i\mathbf{X}_{\kvec 1})^T\text{ } (I_i\mathbf{X}_{\kvec 2})^T\text{ } (I_i\mathbf{X}_{\kvec 3})^T\text{ } (I_i\mathbf{X}_{\kvec 4})^T]$ is the Bogoliubov transformation matrix, the transposed eigenket $\psi'=[b_{1 \kvec}\text{ } b_{2 \kvec}\text{ } b_{3 \kvec}\text{ } b_{4 \kvec}\text{ } b^{\dagger}_{1 -\kvec}\text{ } b^{\dagger}_{2 -\kvec}\text{ } b^{\dagger}_{3 -\kvec}\text{ } b^{\dagger}_{4 -\kvec}]$, 
and $[I_{i}]_{8 \times 8}$ is the exchange (reverse identity) matrix. 

\subsection*{Multiband indirect K--edge RIXS}
We employ a perturbation expansion in the core hole potential to define the RIXS operator as~\cite{nagaoPhysRevB.75.214414,brink0295-5075-80-4-47003}. 
\begin{equation}
\label{eq:bimag}
    \mathcal{O}(\mathbf{\mathbf{q}})=\sum_{i,j}J_{ij}e^{i\mathbf{\mathbf{q}}\cdot\mathbf{\mathbf{R}_i}}\mathbf{S}_i\cdot \mathbf{S}_j,
\end{equation}
where $\mathbf{\mathbf{q}}$ is the X-ray transfer momentum. The RIXS operator $\mathcal{O}(\mathbf{\mathbf{q}})=\sum_{\kvec}\psi^{\dagger}_{\kvec+\mathbf{\mathbf{q}}}\mathcal{O}_{\kvec}\psi_{\kvec}$ after the spin wave theory transformation becomes
\begin{eqnarray}
\mathcal{O}_{\kvec}(\mathbf{q})=
\begin{bmatrix}
M_{A} & M_{B}\\
M_{B} & M_{A} 
\end{bmatrix},
\end{eqnarray}
with
\begin{eqnarray}
M_{A}=
\begin{bmatrix}
A_1(\mathbf{q}) & F_{\kvec}+F_{\kvec+\mathbf{q}} & G_{\kvec}+G_{\kvec+\mathbf{q}} & S_{\kvec}+S_{\kvec+\mathbf{q}}\\
F_{\kvec}^*+F_{\kvec+\mathbf{q}}^* & A_2(\mathbf{q}) & S_{\kvec}+S_{\kvec+\mathbf{q}} & T_{\kvec}+T_{\kvec+\mathbf{q}}\\
G_{\kvec}^*+G_{\kvec+\mathbf{q}}^* & S_{\kvec}^*+S_{\kvec+\mathbf{q}}^* & A_1^*(\mathbf{q}) & F_{\kvec}^*+F_{\kvec+\mathbf{q}}^*\\
S_{\kvec}^*+S_{\kvec+\mathbf{q}}^* & T_{\kvec}^*+T_{\kvec+\mathbf{q}}^* & F_{\kvec}+F_{\kvec+\mathbf{q}} & A_2^*(\mathbf{q})
\end{bmatrix}
\end{eqnarray}
and
\begin{eqnarray}
M_{B}=
\begin{bmatrix}
0 & B_{\kvec}^*+B_{\kvec+\mathbf{q}}^* & C_{\kvec}^*+C_{\kvec+\mathbf{q}}^* & L_{\kvec}^*+L_{\kvec+\mathbf{q}}^*\\
B_{\kvec}+B_{\kvec+\mathbf{q}} & 0 & L_{\kvec}^*+L_{\kvec+\mathbf{q}}^* & D_{\kvec}^*+D_{\kvec+\mathbf{q}}^*\\
C_{\kvec}+C_{\kvec+\mathbf{q}} & L_{\kvec}+L_{\kvec+\mathbf{q}} & 0 & B_{\kvec}+B_{\kvec+\mathbf{q}}\\
L_{\kvec}+L_{\kvec+\mathbf{q}} & D_{\kvec}+D_{\kvec+\mathbf{q}} & B_{\kvec}^*+B_{\kvec+\mathbf{q}}^* & 0
\end{bmatrix}.
\end{eqnarray}
The off diagonal matrix entries are given by $E_{1}=S(-2J_{1}-J_{2}+J_{1}'+2J_{2}'+2J_{3}-J_{3}')$, $F_{\kvec}=J_{1}S e^{i(\kvec\cdot \G_{1,2})}$, $G_{\kvec}=J_{2}S e^{i(\kvec\cdot\G_{1,3})}+J_{3}'Se^{i(\kvec\cdot\G_{1,3}^{\prime\prime})}$, $S_{\kvec}=J_{1}S e^{i(\kvec\cdot\G_{1,4})}$, $B_{\kvec}=J_{2}'S e^{i(\kvec\cdot\G_{2,1}')}+J_{3}Se^{i(\kvec\cdot\G_{2,1}^{\prime\prime})}$, $C_{\kvec}=J_{1}'S e^{i(\kvec\cdot\G_{3,1}')}$, $D_{\kvec}=J_{1}'S e^{i(\kvec\cdot\G_{4,2}')}$, $T_{\kvec}=J_{2}S e^{i(\kvec\cdot\G_{2,4}')}+J_{3}'Se^{i(\kvec\cdot\G_{2,4}^{\prime\prime})}$, and $L_{\kvec}=J_{2}'S e^{i(\kvec\cdot\G_{2,3}')}+J_{3}Se^{i(\kvec\cdot\G_{2,3}^{\prime\prime})}$ where $\G$ denotes the intrablock separation, $\G'$ the NN interblock separation, and $\G^{\prime\prime}$ denotes the n--NN interblock separation. The separation vectors between the spin sites are scaled by the lattice constant. The diagonal coefficients are defined as 
\begin{eqnarray}
A_1(\mathbf{q})&=&E_{1}+S[J_{3}(e^{i2q_{y}}+e^{-i2q_{x}})-J'_{3}e^{i2q_{x}}+J'_{2}(e^{i(q_{x}-q_{y})}+e^{i(q_{x}+q_{y})})\nonumber\\
&&-J_{2}e^{-i(q_{x}+q_{y})}+J'_{1}e^{iq_{y}}-J_{1}(e^{-iq_{x}}+e^{-iq_{y}})],\\
A_2(\mathbf{q})&=&E_{1}+S[J_{3}(e^{-i2q_{y}}+e^{-i2q_{x}})-J'_{3}e^{i2q_{y}}+J'_{2}(e^{i(q_{x}+q_{y})}+e^{i(-q_{x}+q_{y})})\nonumber\\
&&-J_{2}e^{i(q_{x}-q_{y})}+J'_{1}e^{-iq_{x}}-J_{1}(e^{iq_{x}}+e^{-iq_{y}})].
\end{eqnarray}
To obtain the RIXS intensity we utilize the Bogoliubov transformed RIXS operator which is given by
\begin{equation}
    \mathcal{O}'(\mathbf{q})_{\kvec}=\mathbf{X}^{-1}_{\kvec}\mathcal{O}_{\kvec}\mathbf{X}_{\kvec},
\end{equation}
where the above operator can be recast in terms of four matrices as
\begin{eqnarray}
\label{eq:orixsmat1}
    \mathcal{O}'(\mathbf{q})_{\kvec}=
    \begin{bmatrix}
    \mathcal{O}_{11} & \mathcal{O}_{12}\\
    \mathcal{O}_{21} & \mathcal{O}_{22}.\\
    \end{bmatrix}
\end{eqnarray}
The off diagonal matrices of $\mathcal{O}_{12}$ and $\mathcal{O}_{21}$ have entries given by 
\begin{eqnarray}
\label{eq:orixsmat2}
    \mathcal{O}_{12}=\mathcal{O}_{21}=
    \begin{bmatrix}
    W_{11} & W_{12} & W_{13} & W_{14} \\
    W_{21} & W_{22} & W_{23} & W_{24} \\
    W_{31} & W_{32} & W_{33} & W_{34} \\
    W_{41} & W_{42} & W_{43} & W_{44} \\
    \end{bmatrix}.
\end{eqnarray}The individual entries are termed the intra- or inter- band single channels (see Supplementary Fig. S2 and Fig. S3). The multiband indirect K -edge bimagnon RIXS operator constructed from these entries is expressed as 
\begin{equation}
\label{eq:o2mb2}
    \mathcal{O}_{2}(\mathbf{\mathbf{q}})=\sum_{\kvec}\left(\sum_{i,j}W_{ij}b^{\dagger}_{i \kvec+\mathbf{\mathbf{q}}}b^{\dagger}_{j -\kvec}+\sum_{i,j}W^{*}_{ij}b_{i \kvec+\mathbf{\mathbf{q}}}b_{j -\kvec}\right),
\end{equation}
 where the sublattice indices span over $i,j = 1, ..., 4$. We introduce the time-ordered correlation function  
\begin{equation}
    G(\mathbf{\mathbf{q}},\omega)=-i\int^{\infty}_0dte^{i\omega t}\langle i|\mathcal{T}\mathcal{O}_2^{\dagger}(\mathbf{\mathbf{q}},t)\mathcal{O}_2(\mathbf{\mathbf{q}},0)|i\rangle,
\end{equation}
to compute the frequency and momentum dependent RIXS intensity, see equation (\ref{eq:rixsmulb}). The matrix elements of the RIXS operators are taken between $|i\rangle$ and $|f\rangle$ which represent the initial and final states with transfer energy $\omega_{fi}$. The noninteracting LSWT bimagnon propagator is defined as
\begin{equation}
\label{eq:pi0}
    \Pi^{0}_{ij}(\mathbf{\mathbf{q}},t;\kvec,\kvec')=-i\langle 0|\mathcal{T}b_{i,\kvec+\mathbf{\mathbf{q}}}(t)b_{j,-\kvec}(t)b^{\dagger}_{i,\kvec^{'}+\mathbf{\mathbf{q}}}(0)b_{j,-\kvec^{'}}^{\dagger}(0)|0\rangle.
\end{equation}
We can expand the above bimagnon propagator in terms of the one magnon propagators. Since there are four band indices we have four possible combinations given by
\begin{equation}
\label{eq:gfbb}
    G_{b_{i},b_{j}}(\kvec,t)=-i\langle 0|\mathcal{T}b_{i,\kvec}(t)b^{\dagger}_{j,\kvec}(0)|0\rangle,
\end{equation}
where $i,j = 1, 2, 3,$ and $4$ for the $b_{1}$, $b_{2}$, $b_{3}$, and $b_{4}$ magnons, $\mathcal{T}$ is the time ordering operator, and $|0\rangle$ is the ground state. Combining equations (\ref{eq:o2mb2}), (\ref{eq:pi0}), and (\ref{eq:gfbb}) we obtain
\begin{eqnarray}
   \mathsf{S}({\bf q},\omega)&\propto& -\frac{1}{\pi}\Im m\left[\sum\limits_{\kvec,\kvec^{'}}\sum_{ij}|W_{ij}|^2\Pi^{0}_{ij}(\mathbf{\mathbf{q}},\omega;\kvec,\kvec')\right]=-\frac{1}{\pi}\Im m\left[\sum\limits_{\kvec}\sum_{ij}|W_{ij}|^2\frac{1}{\omega-\omega_{i,\kvec+\mathbf{\mathbf{q}}}-\omega_{j-\kvec}+i0^+}\right],
\end{eqnarray}
which is same as equation (\ref{eq:rixsmulb}).

\section*{Data Availability}
All data generated or analysed during this study are included in this published article (and its Supplementary Information files).

\bibliography{chiralbib}
%

\section*{Acknowledgements (not compulsory)}

S.M. and T.D. acknowledge Augusta University  2017 CURS Summer Scholars Program. T.D. acknowledges funding support from Sun Yat-Sen University Grant No. OEMT--2017--KF--06 and Augusta University Scholarly Activity Award. Z. Y. H. and D.X.Y. are supported by NKRDPC-2017YFA0206203, NKRDPC-2018YFA0306001, NSFC-11574404, NSFG-2015A030313176, National Supercomputer Center in Guangzhou, and Leading Talent Program of Guangdong Special Projects. The authors thank Dr. Kenji Ishii for useful discussions on $K$ and $L$ -edge X-ray beamline resolution capabilities.

\section*{Author contributions statement}

T.D. (in consultation with D. X. Y) conceived the idea for the project. S.M. and Z.H. performed the spin wave theory calculations. S.M. calculated the RIXS spectra. T.D. and D.X.Y checked the calculations. T.N. calculated the XAS spectra and the resonance factor contribution to the RIXS spectrum arising from the polarization and Bragg angle dependence. All authors contributed to the analysis and interpretations of the results. All authors contributed to the writing of the manuscript. 

\section*{Additional information}

\textbf{Competing financial interests} The authors declare no competing interests. 


%


\section*{Supplementary Figure 1}
\begin{figure}[H]
\centering
\includegraphics[width=0.75\linewidth]{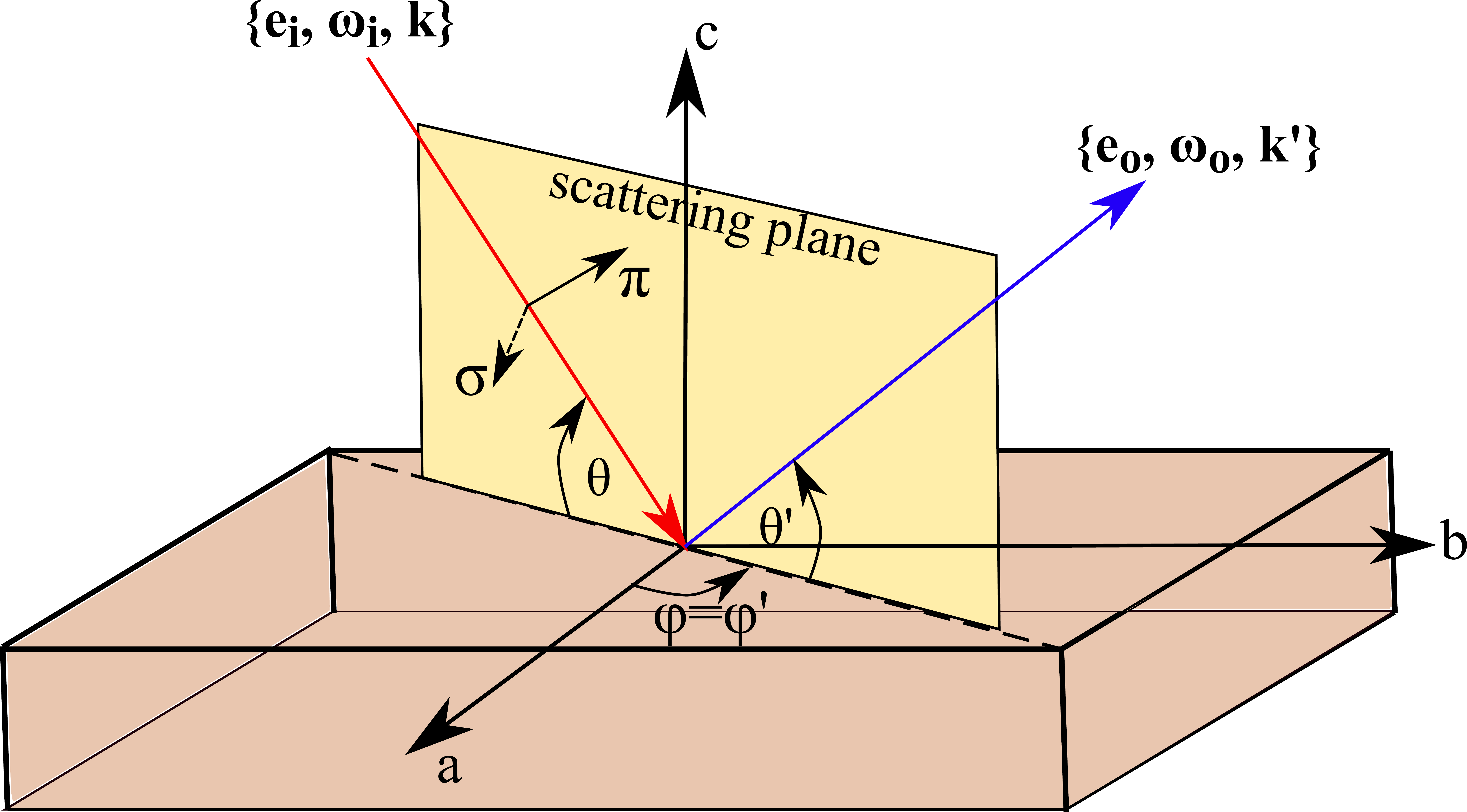}
\caption{Sketch of the scattering geometry displaying the definitions of incoming $\{e_{i},\omega_{i},{\bf k}_{i}\}$ and outgoing $\{e_{o},\omega_{o},{\bf k}_{o} \}$ polarization, energy, and wave vector respectively. The scattering plane (yellow) shows the two possible directions of polarization, in-plane polarization ($\pi$) and the out-of-plane ($\sigma$). The incoming (outgoing) Bragg angles are denoted by $\theta (\theta^{\prime})$. The in-plane azimuthal angles are given by $\phi$ and $\phi^{\prime}$.}
\label{fig:S0}
\end{figure}
\newpage
\section*{Supplementary Figure 2}
With four participating bands there are sixteen channels. These channels are classified as $m-n$, where $m,n =1, 2, 3,$ and $4$. For example, $2-3$ refers to a transition between band number 2 and band number 3. 
\begin{figure}[h]
\centering
\includegraphics[width=\linewidth]{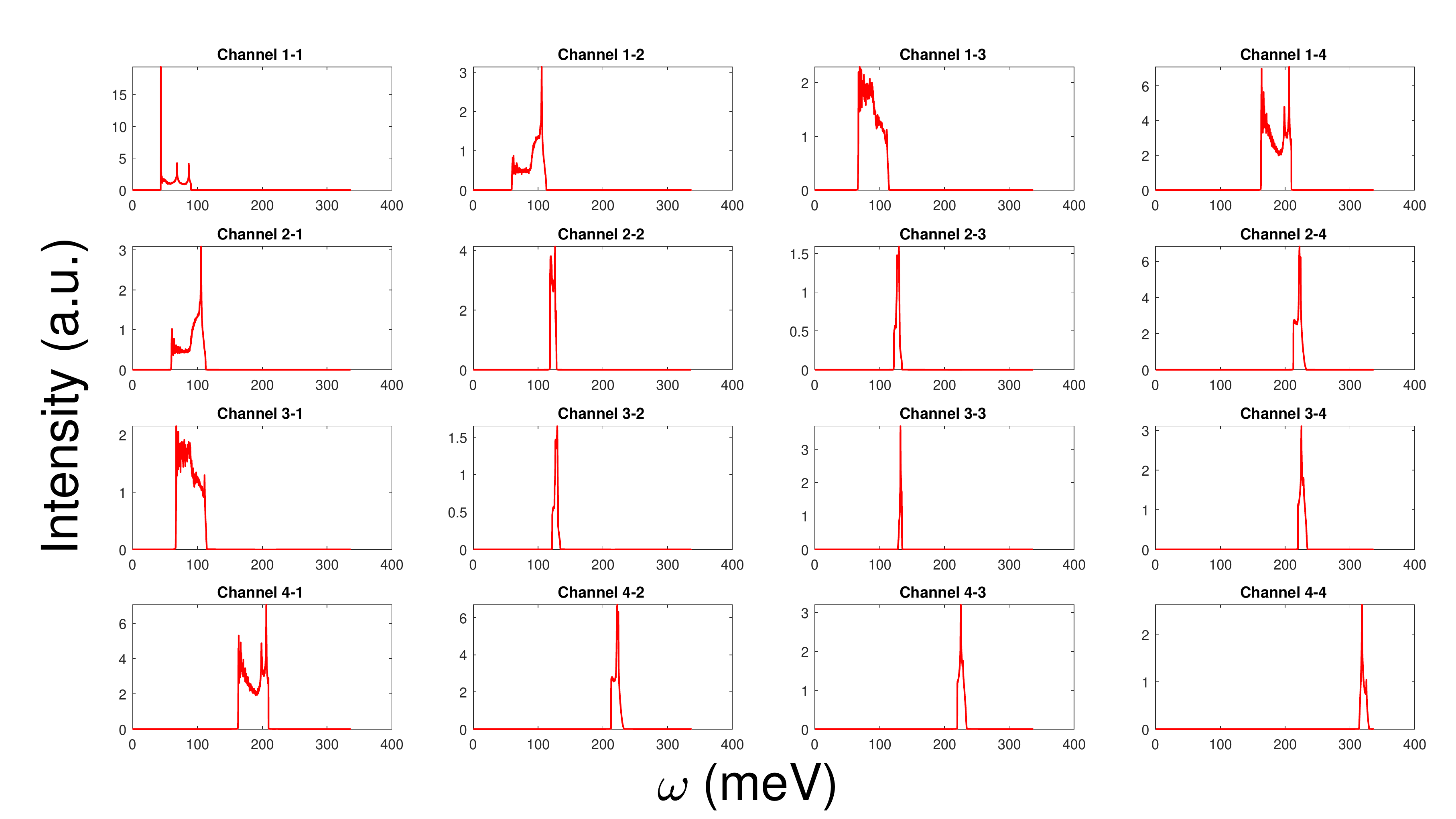}
\caption{RIXS channel responses for the right chiral lattice with the right ordering wave vector. $x-$ axis represents energy in meV. $y-$ axis is normalized intensity.}
\label{fig:S1}
\end{figure}
\newpage
\section*{Supplementary Figure 3}
With four participating bands there are sixteen channels. These channels are classified as $m-n$, where $m,n =1, 2, 3,$ and $4$. For example, $2-3$ refers to a transition between band number 2 and band number 3. 
\begin{figure}[H]
\centering
\includegraphics[width=\linewidth]{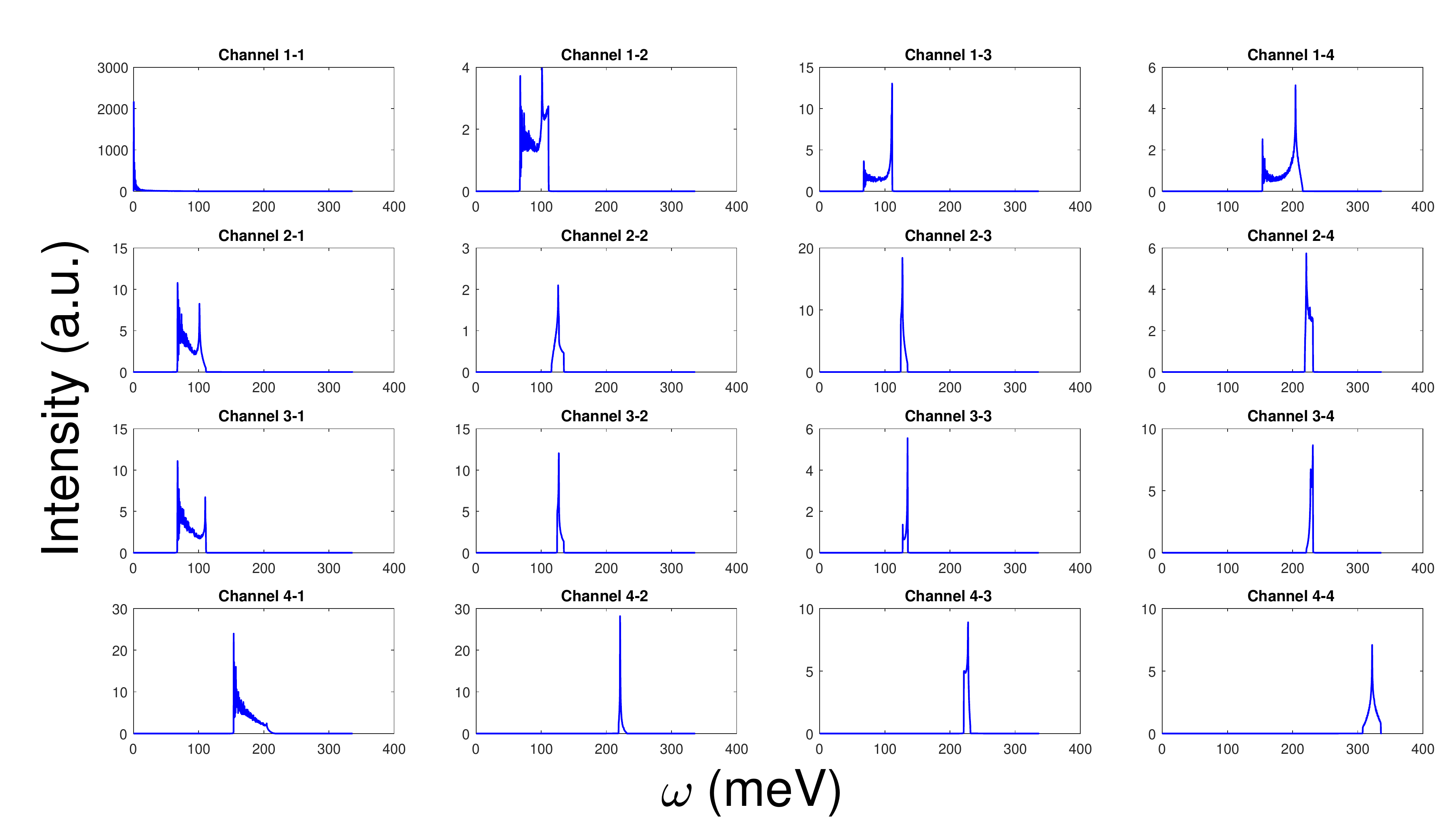}
\caption{RIXS channel responses for the left chiral lattice with the right ordering wave vector. $x-$ axis represents energy in meV. $y-$ axis is normalized intensity. The $i-j$ and $j-i$ channel responses should be symmetric. However, in this case we note a discrepancy which is potentially due to a computational issue, rather than breakdown of any fundamental physical process.}
\label{fig:figS2}
\end{figure}
\newpage
\section*{Supplementary Figure 4}
\begin{figure}[H]
    \centering
    \includegraphics[width=\linewidth]{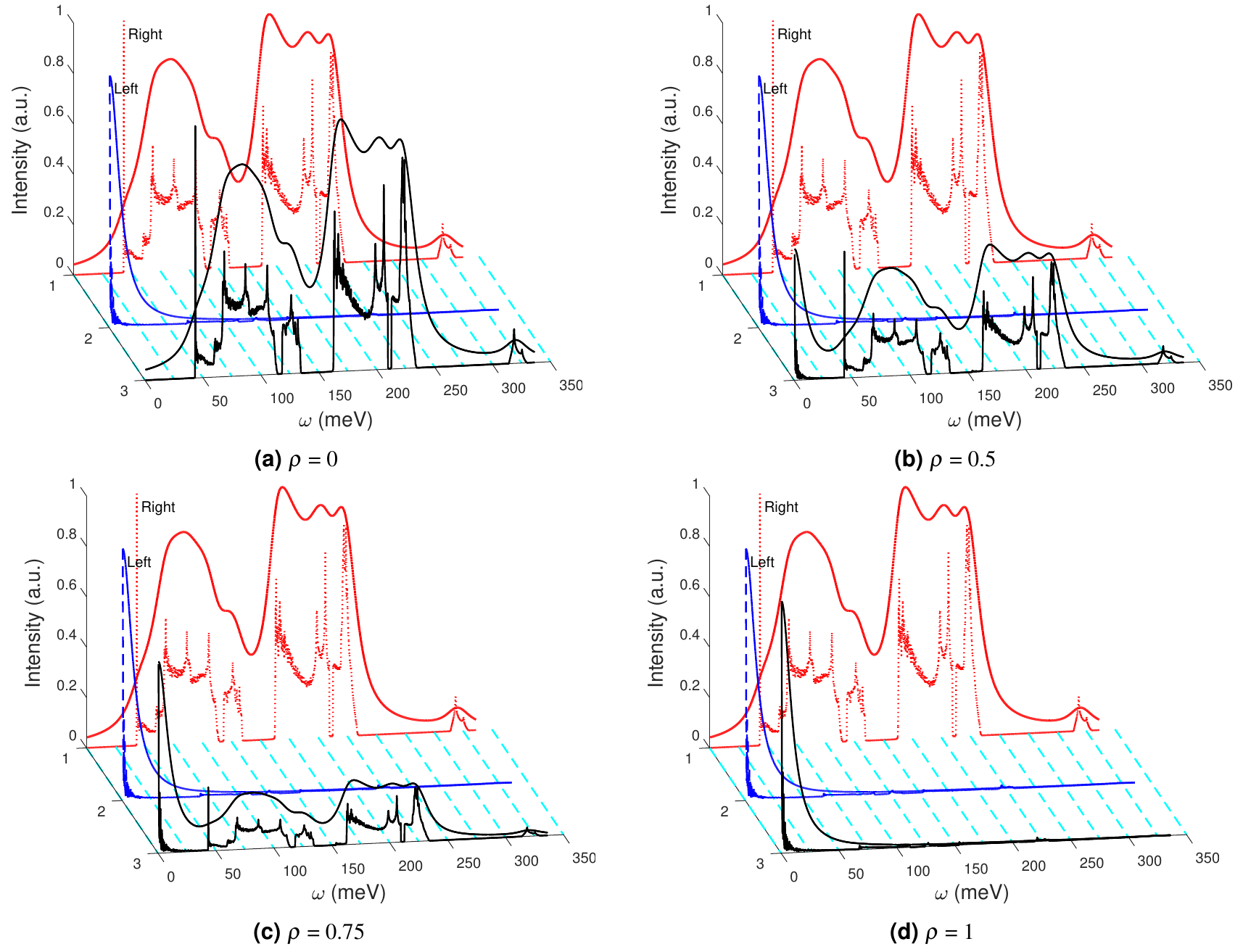}
    ~ 
\caption{Chiral composition mixture response for Flack parameter $\rho$ variation with 10 meV resolution. All responses are for the right chiral ordering wave vector and includes the effects of the resonance factor. (a) $\rho=0.0$ (pure right chiral), (b) $\rho=0.5$ (racemic conglomerate), (c) $\rho=0.75$, and (d) $\rho=1.0$ (pure left chiral).}
\label{fig:RMixROWVmn24full10}
\end{figure}
\newpage
\section*{Supplementary Figure 5}
\begin{figure}[H]
    \centering
    \includegraphics[width=\linewidth]{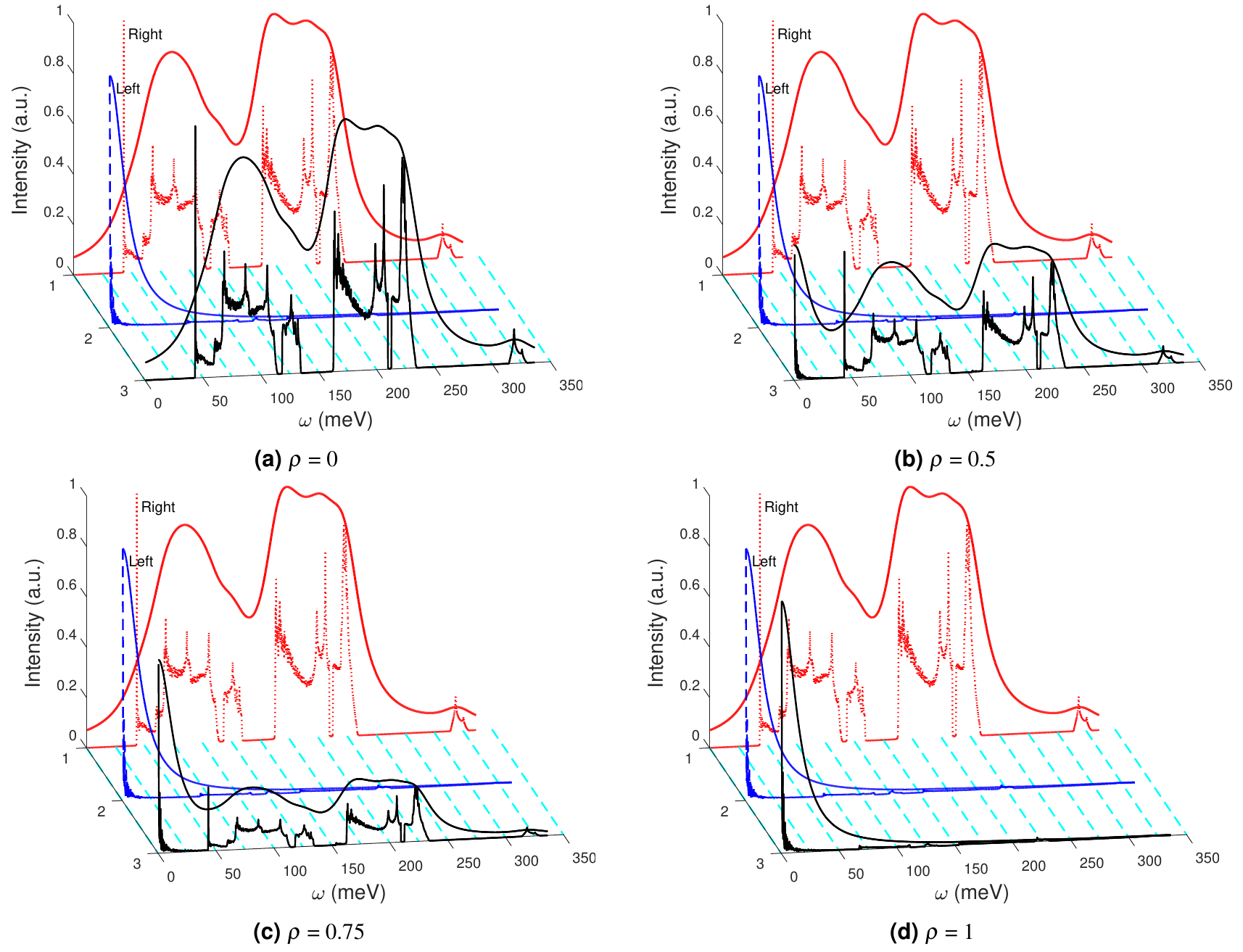}
    ~ 
\caption{Chiral composition mixture response for Flack parameter $\rho$ variation with 15 meV resolution. All responses are for the right chiral ordering wave vector and includes the effects of the resonance factor. (a) $\rho=0.0$ (pure right chiral), (b) $\rho=0.5$ (racemic conglomerate), (c) $\rho=0.75$, and (d) $\rho=1.0$ (pure left chiral).}
\label{fig:RMixROWVmn24full15}
\end{figure}
\newpage
\section*{Supplementary Figure 6}
\begin{figure}[H]
    \centering
    \includegraphics[width=\linewidth]{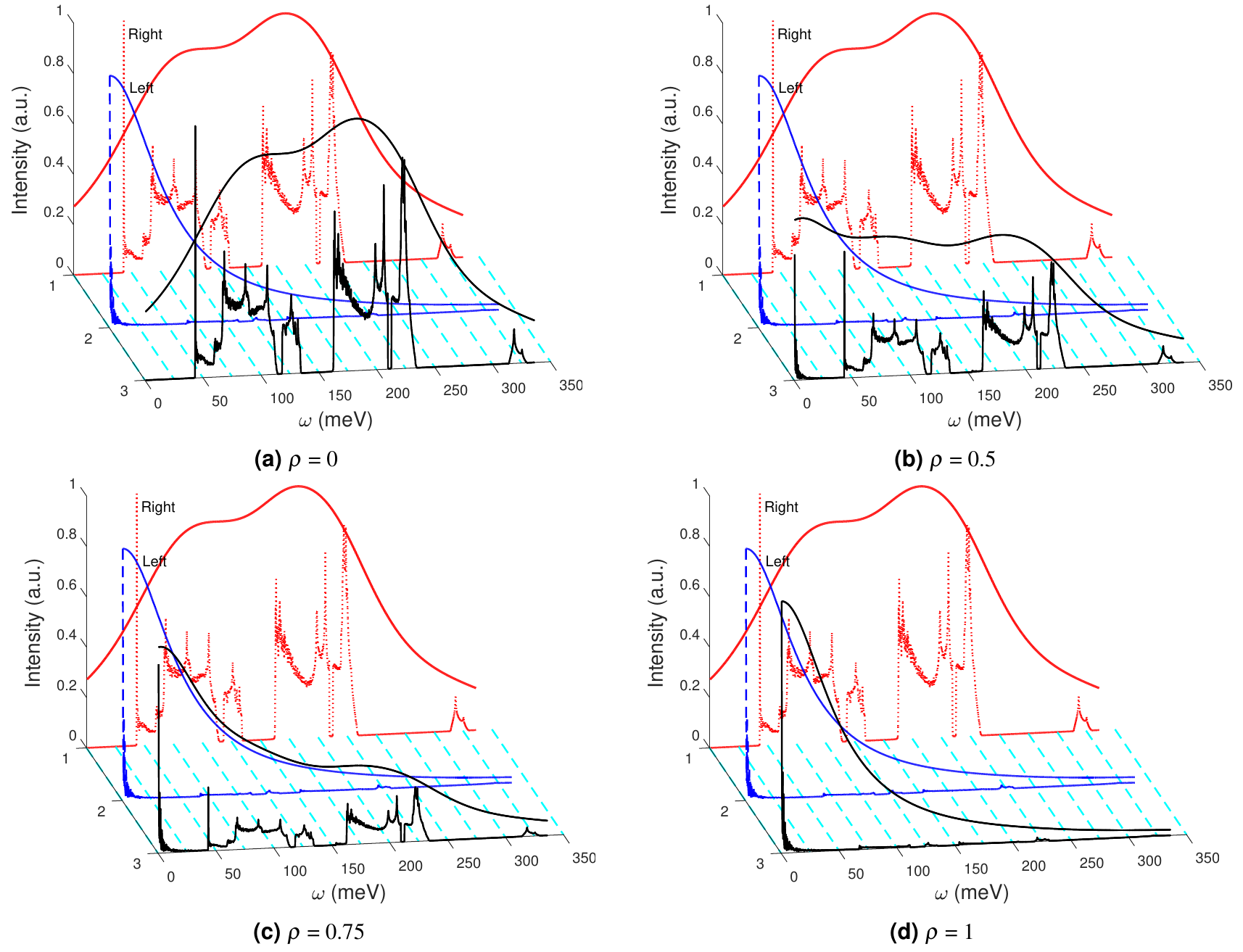}
    ~ 
\caption{Chiral composition mixture response for Flack parameter $\rho$ variation with 50 meV resolution. All responses are for the right chiral ordering wave vector and includes the effects of the resonance factor. (a) $\rho=0.0$ (pure right chiral), (b) $\rho=0.5$ (racemic conglomerate), (c) $\rho=0.75$, and (d) $\rho=1.0$ (pure left chiral).}
\label{fig:RMixROWVmn24full50}
\end{figure}
\newpage
\section*{Supplementary Figure 7}
\begin{figure}[H]
    \centering
    \includegraphics[width=\linewidth]{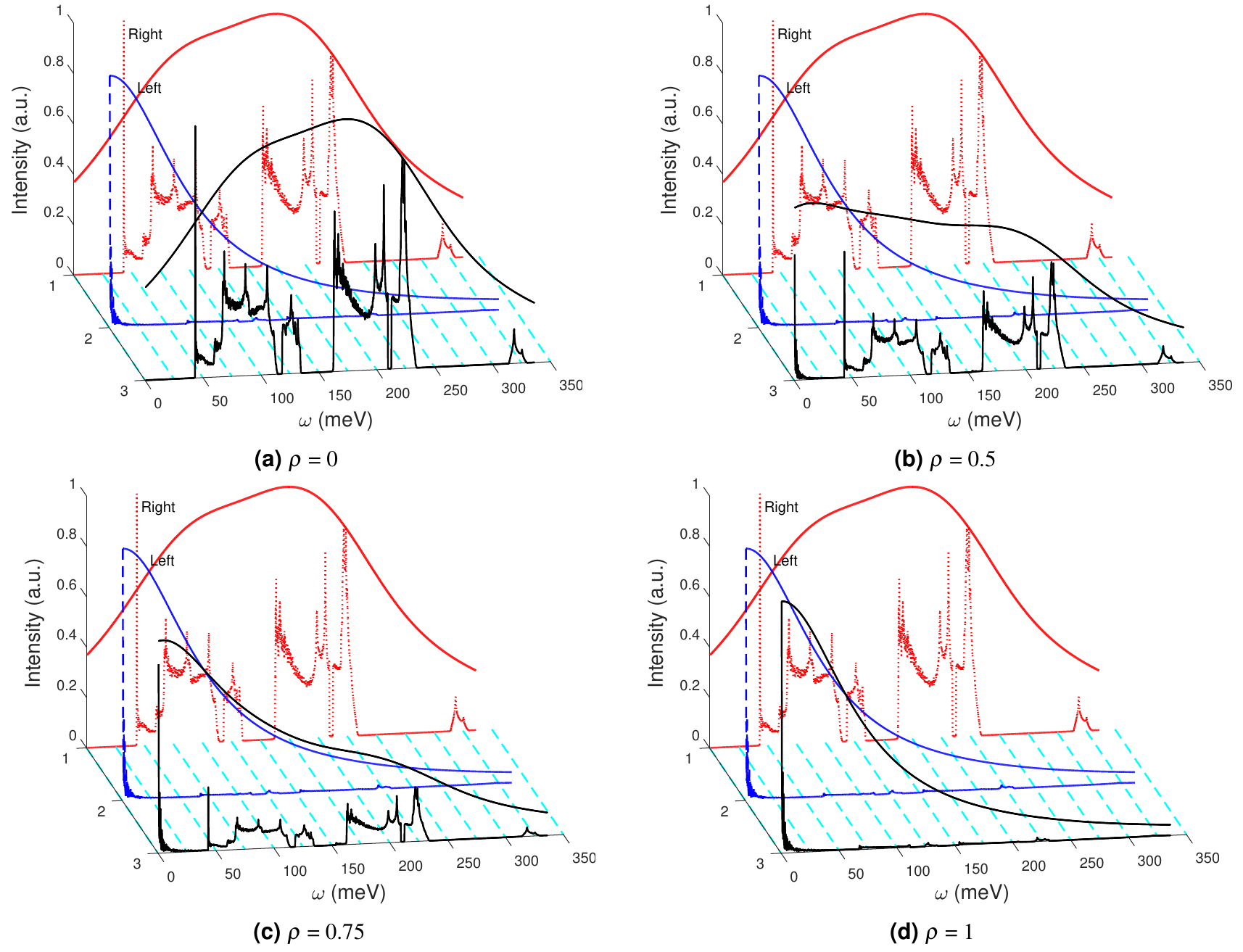}
    ~ 
\caption{Chiral composition mixture response for Flack parameter $\rho$ variation with 70 meV resolution. All responses are for the right chiral ordering wave vector and includes the effects of the resonance factor. (a) $\rho=0.0$ (pure right chiral), (b) $\rho=0.5$ (racemic conglomerate), (c) $\rho=0.75$, and (d) $\rho=1.0$ (pure left chiral).}
\label{fig:RMixROWVmn24full70}
\end{figure}

\end{document}